\documentclass[aps,pra,preprint,floatfix]{revtex4-1}

\usepackage{graphicx}
\usepackage{amsmath}
\usepackage{amssymb}    
\usepackage{bbm}
\usepackage{longtable}
\usepackage{array}

\newcommand{\boldnabla}{\mbox{\boldmath$\nabla$}}
\begin{document}
\title{Numerical solution of the time-independent Dirac equation for diatomic molecules: B-splines without spurious states}

\author{Fran\c{c}ois Fillion-Gourdeau}
\email{filliong@CRM.UMontreal.ca}
\affiliation{Centre de Recherches Math\'{e}matiques, Universit\'{e} de Montr\'{e}al, Montr\'{e}al, Canada, H3T 1J4}

\author{Emmanuel Lorin}
\email{elorin@math.carleton.ca}
\affiliation{School of Mathematics and Statistics, Carleton University, Ottawa, Canada, K1S 5B6}
\altaffiliation[Also at ]{Centre de Recherches Math\'{e}matiques, Universit\'{e} de Montr\'{e}al, Montr\'{e}al, Canada, H3T 1J4}

\author{Andr\'{e} D. Bandrauk}
\email{Andre.Dieter.Bandrauk@USherbrooke.ca}
\affiliation{Laboratoire de chimie th\'{e}orique, Facult\'{e} des Sciences, Universit\'{e} de Sherbrooke, Sherbrooke, Canada, J1K 2R1}
\altaffiliation[Also at ]{Centre de Recherches Math\'{e}matiques, Universit\'{e} de Montr\'{e}al, Montr\'{e}al, Canada, H3T 1J4}

\date{\today}

\begin{abstract}
Two numerical methods are used to evaluate the relativistic spectrum of the two-centre Coulomb problem (for the $\mbox{H}_{2}^{+}$ and $\mbox{Th}_{2}^{179+}$ diatomic molecules) in the fixed nuclei approximation by solving the single particle time-independent Dirac equation. The first one is based on a min-max principle and uses a two-spinor formulation as a starting point. The second one is the Rayleigh-Ritz variational method combined with kinematically balanced basis functions. Both methods use a B-spline basis function expansion. We show that accurate results can be obtained with both methods and that no spurious states appear in the discretization process.
\end{abstract}

\maketitle


\pagestyle{plain}

\section{Introduction}

There has been a strong interest in the solution of the time-independent Dirac equation in the last few decades motivated mostly by new advances in molecular and nuclear physics. More recently, the fields of laser-matter interaction and optics have also considered this equation because new developments have led to experimental facilities reaching laser intensities above $10^{20}$ W$\cdot$cm$^{-2}$ \cite{RevModPhys.78.309}. The mathematical description of an electron subjected to such intense electromagnetic fields necessitates a relativistic treatment \cite{doi:10.1080/09500340210140740,PhysRevA.59.604,PhysRevA.79.043418} and thus, theoretical efforts should be based on the Dirac equation instead of the non-relativistic Schr\"odinger equation. 

It is well-known that finding a solution of the Dirac equation is very challenging because it has an intricate matrix structure. This complicates analytical approaches and closed-form solutions can be found only for highly symmetric systems. For this reason, a numerical treatment is required to study more realistic physical processes occurring in molecules or heavy ion collisions. However, the development of an accurate numerical approach is also difficult because the Dirac spectrum is not bounded from below (and above). This precludes the ``naive'' generalization of well-known methods used to solve the Schr\"odinger equation such as the Galerkin or Rayleigh-Ritz methods. These are based on minimization principles and thus, work rigorously only if the spectrum of the differential operator has a lower (or upper) bound. Trying to solve the Dirac equation with these methods can thus be very dangerous as the Dirac spectrum can be altered by negative energy contributions. This problem is named the ``variational collapse'' and leads to the appearance of spurious states in the approximated spectrum. The spurious states are eigenvalues which do not belong to the spectrum of the continuous operator and which appear in the discretization process. More precisely, let $\lambda$ be an eigenvalue of the Dirac operator $\hat{H}$ in the mass gap $(-mc^{2},mc^{2})$ (corresponding to bound states) and $\sigma_{\hat{H}}$ be its point spectrum in the mass gap, that is the set of all $\lambda$'s. Numerically, we approximate the operator $\hat{H}$ by $\hat{H}_{n}$ such that $\lim_{n \rightarrow \infty}\hat{H}_{n}=\hat{H}$ (here, $n$ is the dimension of the subspace on which the operator is projected or, loosely speaking, the dimension of the Dirac operator matrix once the problem is discretized). The discretized Dirac operator $\hat{H}_{n}$ has eigenvalues given by $\lambda_{n} \in \sigma_{\hat{H}_{n}}$. The set of spurious states $\sigma^{\rm s}_{\hat{H}_{n}} \subset \sigma_{\hat{H}_{n}}$ is defined by the set of all eigenvalues in $\sigma_{\hat{H}_{n}}$ for which $ \lim_{n \rightarrow \infty} \lambda_{n} \notin \sigma_{\hat{H}}$.     

There have been many successful attempts in the literature to circumvent variational collapse by adapting minimization techniques and basis set expansions to the Dirac operator \cite{10.1063/1.447865,PhysRevA.23.2093,PhysRevA.25.1230,PhysRevA.40.5559,Desclaux2003453,bandrauk_book,PhysRevA.62.022508,QUA:QUA560250112,PhysRevA.28.3092,grant2006relativistic}. Usually, these techniques can be classified into one of three main categories \cite{QUA:QUA560250112}: modification of basis functions, utilization of an operator that has a lower bound but the same spectrum as the Dirac operator and transformation of the Dirac operator. The methods utilized in this work fall into the first and third categories; they are the Rayleigh-Ritz method with kinematically balanced basis functions \cite{PhysRevA.62.022508,0022-3700-19-20-003,1402-4896-36-3-013,grant2006relativistic} and the variational method based on a min-max principle \cite{PhysRevLett.85.4020,Dolbeault2003}.

The discretization of the Dirac equation in these two cases proceeds by the utilization of a basis set expansion which allows, in principle, a very good accuracy. These techniques have been exploited extensively in the non-relativistic case to solve the time-independent Schr\"odinger equation \cite{10.1063/1.1680025} and allow making accurate predictions for the non-relativistic spectra of molecules and for other observables. In this work, a B-spline set of basis functions will be utilized. This choice is motivated by some interesting properties of B-splines: they have compact support (leading to sparse matrix structures), they are very flexible in terms of element size and continuity conditions (both are determined by their knot vector), they are positive definite and finally, they are linearly independent (they form a complete basis). For these reasons, B-splines have also been widely applied to the non-relativistic (see \cite{0034-4885-64-12-205} for a review on the use of B-spline in molecular physics) and to the relativistic \cite{PhysRevA.37.307,PhysRevLett.93.130405,Dolbeault2003,JPSJ.75.114301,0953-4075-43-23-235207} cases. Here, we combine B-spline basis functions with the two numerical schemes described previously. 

We use these methods to investigate two particular systems: the diatomic molecule $\mbox{H}_{2}^{+}$ and the quasi-molecule $\mbox{Th}_{2}^{179+}$. This is accomplished by computing the relativistic spectrum of the two-centre Coulomb problem in the fixed nuclei approximation. The rationale for studying these systems is twofold. First, they are physically relevant in many fields of physics. The molecule $\mbox{H}_{2}^{+}$ is very important in chemical physics and its relativistic corrections, albeit very small, have been the subject of many studies \cite{1402-4896-36-3-004,Laaksonen1984485,Dage1994469,Kullie1999307,Kullie2004215}. On the other hand, the quasi-molecule $\mbox{Th}_{2}^{179+}$ is not stable and dissociates rapidly. However, it is pertinent in heavy ion collisions at intermediate energy, where processes such as charge transfer and electron-positron pair production are investigated \cite{PhysRevA.24.103,PhysRevA.56.2806}, and in high intensity laser-matter interaction, where pair production and Quantum Electrodynamics (QED) processes could be enhanced in the presence of heavy nuclei \cite{PhysRevA.73.062106}. 

The second reason to look at these systems is that their ground state eigenvalue has already been computed and thus, the numerical results obtained from our analysis can be compared to results from the literature. More precisely, they have been studied using different analytical \cite{Muller19735,Greiner:1987} and numerical approaches such as the Rayleigh-Ritz scheme \cite{0305-4470-16-9-025,Laaksonen1984485,PhysRevA.48.2700,Kullie2004215,Kullie1999307,0953-4075-43-23-235207}, the variational scheme based on the min-max principle \cite{Dolbeault2003} and finite difference methods \cite{1402-4896-36-3-004,Dage1994469}. Very accurate results for the ground state of diatomic molecules were obtained in these analyses. However, the whole spectrum is rarely discussed (an exception to this is found in \cite{0953-4075-37-4-016,0953-4075-38-16-008}) and some of these methods (especially those based on the ``naive'' Rayleigh-Ritz method) could potentially lead to the appearance of spurious states. For instance, this problem was discussed in \cite{0953-4075-43-23-235207} and a technique for identifying these artifacts was described. However, this can be cumbersome when one is interested in sums over intermediate states such as those required in radiative QED corrections. For these calculations, it is certainly more efficient to have a numerical scheme free from spurious states from the outset. 

In this work, we are presenting and comparing two numerical methods that use a B-spline basis set expansion to compute the relativistic spectrum of the two-centre problem. The first one is the min-max variational method. The second one is the Rayleigh-Ritz method combined with kinematically balanced basis function. In Section \ref{sec:num_meth}, the variational formulation of both numerical methods is presented and the choice of basis functions is described. The numerical results are displayed in Section \ref{sec:res} where some values for the spectra of diatomic molecules are shown along with a discussion of spurious states. The conclusion is found in Section \ref{sec:conclusion}.

%

\section{Numerical methods}
\label{sec:num_meth}

The Dirac equation describes the relativistic dynamics of spin-$\frac{1}{2}$ particles (fermions) like the electron. In this work, we consider specifically the single particle static Dirac equation without vector potential given by \footnote{All the calculations will be performed in atomic units (a.u.) where $m=1$, $\hbar =1$ and $c=1/\alpha $ where we take $\alpha \approx 1/137.035999679$ as the fine structure constant. In all the equations however, we are keeping the mass explicitly, allowing to switch easily from atomic to natural units.}
\begin{eqnarray}
\label{eq:dirac}
\hat{H} \psi (x) = E \psi(x) \;\;\mbox{with} \;\; \hat{H} \equiv  c \boldsymbol{\alpha} \cdot \mathbf{p} +  mc^{2} \beta + V(x) \mathbb{I}_{4} ,
\end{eqnarray}
where $\hat{H}$ is the Hamiltonian operator, $\mathbf{p} = -i\boldnabla$ is the momentum operator, $c$ is the light velocity, $m$ is the electron mass, $E$ is the electron energy and $\psi \in L^{2}(\mathbb{R}^{3},\mathbb{C}^{4})$ is a four-spinor. The matrix structure is given by $\boldsymbol{\alpha}$ and $\beta$ which are four by four matrices given by
\begin{eqnarray}
 \alpha_{i} = \left[
\begin{array}{cc}
 0 & \sigma_{i} \\
\sigma_{i} & 0
\end{array} \right]
\; \mbox{and} \;
 \beta = \left[
\begin{array}{cc}
 \mathbb{I}_{2} & 0 \\
0 & -\mathbb{I}_{2}
\end{array} \right],
\end{eqnarray}
where $\sigma_{i}$ are the usual Pauli matrices. The latter are
\begin{eqnarray}
 \sigma_{x} = \left[
\begin{array}{cc}
 0 & 1 \\
1 & 0
\end{array} \right]
\; \mbox{,} \;
\sigma_{y} = \left[
\begin{array}{cc}
 0 & -i \\
i & 0
\end{array} \right]
\; \mbox{and} \;
 \sigma_{z} = \left[
\begin{array}{cc}
 1 & 0 \\
0 & -1
\end{array} \right].
\end{eqnarray}
Written in this way, the Dirac equation is in the Dirac representation. This equation gives a consistent description of single bound electrons within the fixed nuclei approximation, i.e. when the effect of the nuclei is included in the potential term $V$ and the nuclei are fixed in space. This is valid when the masses of the nuclei are much larger than the mass of the electron, which will always be the case for the systems considered in this study.

The main goal of this work is to calculate approximate solutions of (\ref{eq:dirac}). To achieve this, it is convenient to write the four-spinor as $\psi (x) \equiv \left[ \phi(x) , \chi(x) \right]^{\rm T}$ where $\phi(x)$ and $\chi(x)$ are two bi-spinors called the large and small components, respectively. The Dirac equation then becomes
\begin{eqnarray}
\label{eq:dir_mat}
 \left[
\begin{array}{cc}
 V(x) + mc^{2} & \hat{R} \\
\hat{R} & V(x) - mc^{2}
\end{array} \right]
 \left[
\begin{array}{c}
 \phi(x) \\
 \chi(x)
\end{array} \right]
=
E
\left[
\begin{array}{c}
 \phi(x) \\
 \chi(x)
\end{array} \right],
\end{eqnarray}
where we defined $\hat{R} = -ic \boldsymbol{\sigma} \cdot \boldnabla$. This last equation is the common starting point for the numerical methods that follow. As will be seen later, it is also handy to decompose the latter into two coupled equations as 
\begin{eqnarray}
 \hat{R} \chi(x) &=& [E-mc^{2} - V(x)] \phi(x) , \\
\hat{R} \phi(x) &=& [E+mc^{2} - V(x)] \chi(x) .
\end{eqnarray}
The small component can then be written in terms of the large component, yielding
\begin{eqnarray}
\label{eq:sm_comp}
\chi(x) = \frac{\hat{R}}{E+mc^{2} - V(x)}  \phi(x) .
\end{eqnarray}
By substitution, we get
\begin{eqnarray}
\label{eq:large_comp}
 \hat{R} \left[ \frac{\hat{R} \phi(x)}{E+mc^{2}-V(x)} \right] = [E-mc^{2}-V(x)] \phi(x) ,
\end{eqnarray}
for the large component, which belongs to $\phi \in L^{2}(\mathbb{R}^{3},\mathbb{C}^{2})$. Note here that the latter procedure can also be implemented as a Foldy-Wouthuysen transformation \cite{PhysRev.78.29}. These two relations will be important for the analysis that follows.

For a diatomic molecule, the case considered in this study, the static potential is
\begin{equation}
\label{eq:potential}
 V(x) = - \frac{Z_{1}}{|\mathbf{x} + R\mathbf{\hat{z}}|} - \frac{Z_{2}}{|\mathbf{x} - R\mathbf{\hat{z}}|}.
\end{equation}
where $Z_{1,2}$ are the atomic electric charges, $2R$ is the inter-atomic distance and $\hat{z}$ is a unit vector in the z-coordinate direction. It represents the static Coulomb interaction of two point-like nuclei with an electron in the fixed nuclei approximation. This potential is axially symmetric so the number of dimensions can be reduced by one: the azimuthal coordinate dependence can be treated analytically by factorization. Thus, the four-spinor in cylindrical symmetry reads \cite{0305-4470-16-9-024,Kullie2004215} 
\begin{eqnarray}
\label{eq:ansatz}
 \psi (x) = \left[
\begin{array}{c}
 \phi(\xi,\eta,\theta) \\
\chi (\xi,\eta,\theta)
\end{array}
 \right] = \left[
\begin{array}{c}
 \phi_{1}(\xi,\eta) e^{i(j_{z} - 1/2)\theta} \\
\phi_{2}(\xi,\eta) e^{i(j_{z} + 1/2)\theta} \\
i\chi_{1}(\xi,\eta) e^{i(j_{z} - 1/2)\theta} \\
i\chi_{2}(\xi,\eta) e^{i(j_{z} + 1/2)\theta}
\end{array} 
\right] . 
\end{eqnarray}
where $j_{z}$ is the angular momentum projection on the z-axis (it can take one of the values $j_{z} = ...,-\frac{5}{2},-\frac{3}{2},-\frac{1}{2},\frac{1}{2},\frac{3}{2},\frac{5}{2},...$) and $\eta,\xi$ are prolate spheroidal coordinates. This choice of coordinate system is very convenient for the numerical implementation because the Coulomb singularities are situated on the domain boundaries. Also, it has already been utilized in accurate evaluation of the diatomic ground state energy in the relativistic \cite{Kullie1999307,Kullie2004215} and non-relativistic \cite{PhysRevA.10.51,becke:6037,Kobus1996346} cases. For these reasons, these coordinates will be used throughout this work even though it was argued in \cite{0953-4075-43-23-235207} that Cassini coordinates could provide slightly more accurate results. The prolate spheroidal coordinates are defined as
\begin{eqnarray}
 x &=& R \left[ (\xi^{2} - 1)(1-\eta^{2}) \right]^{\frac{1}{2}} \cos \theta , \\ 
 y &=& R \left[ (\xi^{2} - 1)(1-\eta^{2}) \right]^{\frac{1}{2}} \sin \theta , \\ 
 z &=& R \xi \eta ,
\end{eqnarray}
where $\xi \in [1,\infty)$, $\eta \in [-1,1]$ and $\theta = [0,2\pi]$ (azimuthal angle). The Coulomb potential in these coordinates becomes
\begin{eqnarray}
 V(\xi,\eta) = -\frac{Z_{1}}{R(\xi+\eta)} -\frac{Z_{2}}{R(\xi-\eta)}.
\end{eqnarray}
We can now start discussing the numerical methods utilized in this work to calculate the spectrum of diatomic molecules.


\subsection{Min-max method}

The first method described in this work was developed in \cite{Dolbeault2000208,PhysRevLett.85.4020,dolbeault_2000,Dolbeault20041,Dolbeault2003} and is a weak formulation for operators with gaps in their spectrum. The main idea is to find the critical points of the Rayleigh-Ritz coefficient by using a min-max principle. More precisely, it was rigorously proven that the sequence of values defined by \cite{Dolbeault2000208}
\begin{eqnarray}
 \lambda_{k} = \inf_{
\begin{tabular}{c}
\footnotesize{ $G$ subspace of  $F_{+}$ }\\
\footnotesize{$\dim G=k$ }
\end{tabular} 
}
\sup_{\psi \in (G \oplus F_{-}) \backslash \{0\}} \frac{\langle \psi | \hat{H}| \psi \rangle}{\langle \psi|\psi \rangle} ,
\end{eqnarray}
give the actual eigenvalues of $\hat{H}$ in the mass gap, if certain assumptions are fulfilled (for instance, the first eigenvalue should obey $\lambda_{1}>-mc^{2}$ and the potential $V$ should be Coulomb-like). Here, $F_{+,-}$ are two well-defined orthogonal subspaces of $F \subset L^{2}(\mathbb{R}^{3},\mathbb{C}^{4})$. This result is very general and can be applied to any space decompositions $F_{+} \oplus F_{-}$. For practical purpose, it is convenient to use one that splits the large and small components of the Dirac equation as \cite{Dolbeault2000208}
\begin{eqnarray}
F_{+} = L^{2}(\mathbb{R}^{3},\mathbb{C}^{2}) \otimes \left\{ \left(
\begin{array}{c}
0 \\
0
\end{array}
 \right)\right\} \;,\;
 F_{-} =  \left\{ \left(
\begin{array}{c}
0 \\
0
\end{array}
 \right)\right\} \otimes L^{2}(\mathbb{R}^{3},\mathbb{C}^{2}) .
\end{eqnarray}
In this setting, the maximization is performed exactly by the relation (\ref{eq:sm_comp}) which relates the large and small components \cite{Dolbeault2000208}. Then, only the minimization remains and this step is carried out numerically. 

The latter can be performed by minimizing the energy over any couple $(E,\phi)$ obeying the functional equation given by
\begin{eqnarray}
A[E,\phi] \equiv  \int d^{3}x \left[  \left( \frac{|\hat{R} \phi|^{2}}{E+mc^{2}-V} \right) - [E-mc^{2}-V] |\phi|^{2} \right] =0 .
\label{eq:weak_form}
\end{eqnarray}
This functional equation is obtained from (\ref{eq:sm_comp}) and (\ref{eq:large_comp}) by multiplying the latter by $\phi^{\dagger}$ on the left, by integrating on space (using integration by parts) and by using the divergence theorem. Also, the wave function should vanish faster than $\sim \frac{1}{r^{2}}$ at infinity. This is the case when $V$ is a Coulomb-like potential because the corresponding wave function vanishes as $\phi \sim e^{-r}$ when $r \rightarrow \infty$ \cite{Itzykson:1980rh}. 

Therefore, the last equation gives a realization of the min-max principle. Moreover, it is shown from the preceding procedure that it predicts the same spectrum in the mass gap $(-mc^{2},mc^{2})$ as that of the Dirac equation \cite{Dolbeault2000208,PhysRevLett.85.4020,dolbeault_2000,Dolbeault20041,Dolbeault2003}. This formulation also allows discretization schemes that use a basis function discretization without ``variational collapse'': the calculated energy spectrum is bounded in the mass gap and does not fall into the negative energy continuum. Finally, spurious states does not appear in the calculated spectrum without adding any additional conditions \cite{PhysRevLett.85.4020}, making for a very robust numerical method. In the following, we describe the discretization of (\ref{eq:weak_form}) by using a set of basis functions. 

\subsubsection{Basis set expansion}

The discretization of (\ref{eq:weak_form}) with the potential in (\ref{eq:potential}) proceeds by expanding the wave function over a set of basis functions. Thus, the bi-spinor can be written as
\begin{eqnarray}
\label{eq:basis_1}
 \phi_{1,2}(\xi,\eta) &=& \sum_{n=1}^{N} a_{n}^{(1,2)}B^{(1,2)}_{n}(\xi,\eta) ,
\end{eqnarray}
where $a_{n}^{(1,2)}$ are the basis expansion coefficients and $B^{(1,2)}_{n}(\xi,\eta)$ are the basis functions (to be defined later), for components 1 and 2, respectively.

The explicit expression of (\ref{eq:weak_form}) in discretized form depends on the potential considered, on the coordinate choice and is a complicated functional of basis functions (some examples can be found in \cite{Dolbeault2003,PhysRevLett.85.4020}). The equation $A[E,\phi]=0$, once discretized, generally has the form 
%
\begin{equation}
\label{eq:non_lin_eig}
 \sum_{i,j=1}^{2N} a^{(1,2)}_{k,i} A_{ij}(E)a^{(1,2)}_{k,j} = \Lambda_{k}(E) ,
\end{equation}
where $\mathbf{A}(E)$ is now a matrix, $\Lambda_{k}(E)$ is its $k$'th eigenvalue and  
\begin{equation}
 a^{(1,2)}_{k,i} = \left\{
\begin{array}{ccc}
a^{(1)}_{k,i} & \mbox{for} & i \leq N \\
a^{(2)}_{k,i-N} & \mbox{for} & i > N 
\end{array}
\right. ,
\end{equation}
its eigenvector. For cylindrically symmetric systems expressed in prolate spheroidal coordinates, the case considered in this study, $\mathbf{A}(E)$ becomes a $2N \times 2N$ matrix having the following structure:
\begin{eqnarray}
\mathbf{A}(E) = \left[
\begin{array}{cc}
\mathbf{A}_{11}(E) & \mathbf{A}_{12}(E) \\
\mathbf{A}_{12}^{\rm T}(E)& \mathbf{A}_{22}(E)
\end{array}
\right] ,
\end{eqnarray}
where  $\mathbf{A}_{11},\mathbf{A}_{22}$ and $\mathbf{A}_{12}$ are $N\times N$ matrices for which explicit expressions can be found in Appendix \ref{app:explicit_minmax}.

Therefore, solving the non-linear  ($\mathbf{A}(E)$ depends on the energy $E$) eigenvalue problem (\ref{eq:non_lin_eig}) gives an approximation of the energy $E$ and eigenfunctions: the energy of the $k$'th bound state is a solution of $\Lambda_{k}(E)=0$ where $\Lambda_{k}$ is the $k$'th eigenvalue of $\mathbf{A}(E)$ while wave function coefficients in the basis expansion are the $\mathbf{A}(E)$ matrix eigenvector coefficients. It can easily be demonstrated that $\Lambda_{k}(E)$ are monotonically decreasing functions \cite{Dolbeault2003} which implies that they have only one root, i.e. only one value of $E=E_{\rm root}$ for which $\Lambda_{k}(E_{\rm root})=0$. This problem can thus be solved by iteration or any other root finding algorithm.

\subsection{Rayleigh-Ritz method}

The Rayleigh-Ritz method is well-known and has been studied extensively for both the relativistic and non-relativistic cases (see \cite{PhysRevA.62.022508,grant2006relativistic} for instance). Starting from (\ref{eq:dir_mat}), we can multiply by $\psi^{\dagger}$ on the left and integrate on space to get another functional equation given by
\begin{eqnarray}
\label{eq:ray_ritz}
 \int d^{3}x  \left\{ \left[ mc^{2} + V \right] |\phi|^{2} + (  \phi | \hat{R} \chi ) + ( \chi | \hat{R} \phi ) + [V - mc^{2}] |\chi|^{2} \right\}= \nonumber \\
  E \int d^{3}x  \left\{ |\phi|^{2} + |\chi|^{2} \right\} ,
\end{eqnarray}
which is just an explicit way of writing the well-known Rayleigh-Ritz functional equation $\bar{H}=\langle \psi | \hat{H} |\psi \rangle / \langle \psi | \psi \rangle$. The notation $(\cdot|\cdot)$ stands for the Hermitian inner product. In the following, we define two operators $C$ and $S$ by
\begin{eqnarray}
 C[\psi] &=& \int d^{3}x \left\{ \left[ mc^{2} + V \right] |\phi|^{2} + ( \phi|\hat{R}\chi) + (\chi|\hat{R} \phi) + [V - mc^{2}] |\chi|^{2} \right\}, \\
S[\psi] &=& \int d^{3}x \left\{ |\phi|^{2} + |\chi|^{2} \right\} .
\end{eqnarray}
A numerical scheme can be developed from these equations by discretizing the wave function over a set of basis functions. 

For a bounded operator (like the Schr\"odinger operator), the best estimate for the eigenpairs is obtained by a minimization procedure because the Rayleigh-Ritz quotient forms an upper bound for the eigenenergy $E$ (if the spectrum is bounded from below). Moreover, it can be shown that the minimum of the quotient converges towards the exact eigenpair as the number of basis function $N \rightarrow \infty$ (see \cite{PhysRevA.62.022508} and references therein). If the operator is not bounded, as in the case of the Dirac equation, the quantity $\bar{H}$ does not necessarily form an upper bound, although it is still a stationary point (as seen in the previous section, the eigenvalues can actually be characterized by a min-max principle). For this reason, the convergence of this approach is not guaranteed because the stationary point is a saddle point, and spurious states may appear. Therefore, a modification of the method is required to improve the convergence. The strategy we are using in this paper is the Kinematically Balanced Basis functions (KBBF) described in the next section.



\subsubsection{Basis set expansion}

The discretization of (\ref{eq:ray_ritz}) is very similar to the one in the last section and proceeds by expanding the wave function over a set of basis functions. In this case, one writes the bi-spinors as
\begin{eqnarray}
\label{eq:basis_2}
 \phi_{1,2}(\xi,\eta) &=& \sum_{n=1}^{N} a_{n}^{(1,2)}B^{(1,2)}_{n}(\xi,\eta) ,\\
\label{eq:basis_3}
\chi_{1,2}(\xi,\eta) &=& \sum_{n=1}^{N} c_{n}^{(1,2)} X^{(1,2)}_{n}(\xi,\eta) ,
\end{eqnarray}
where $a_{n}^{(1,2)},c_{n}^{(1,2)}$ are the basis expansion coefficients and $B^{(1,2)}_{n}(\xi,\eta),X^{(1,2)}_{n}(\xi,\eta)$ are the basis functions (to be defined later), for components 1 and 2, respectively. In the naive Rayleigh-Ritz method, the basis functions for both spinors are the same, that is $X^{(1,2)}_{n}(\xi,\eta) = B^{(1,2)}_{n}(\xi,\eta)$. Substituting (\ref{eq:basis_2}) and (\ref{eq:basis_3}) in (\ref{eq:ray_ritz}), we obtain a generalized eigenvalue problem in the form of
\begin{eqnarray}
\label{eq:gen_eig}
 \mathbf{C} \mathbf{a} = E \mathbf{S} \mathbf{a} ,
\end{eqnarray}
where the generalized eigenvector $\mathbf{a} = [ a^{(1)}_{1},...,a^{(1)}_{n},a^{(2)}_{1},...,a^{(2)}_{n},c^{(1)}_{1},...,c^{(1)}_{n},c^{(2)}_{1},...,c^{(2)}_{n} ]$ contains the basis function expansion coefficients. Finding the solution of this last equation corresponds to an extremization on the trial function parameters ($a^{(1,2)}_{i}$ and $c^{(1,2)}_{i}$) and yields an approximation of the eigenenergies and eigenfunctions. The functionals $C[\psi]$ and $S[\psi]$ become $4N \times 4N$ matrices having the following general structure
 \begin{eqnarray}
  \mathbf{C} &=& \left[
 \begin{array}{cccc}
  \mathbf{C}^{(1)}_{11} &0&\mathbf{C}^{(3)}_{11}&\mathbf{C}^{(3)}_{12} \\
 0 & \mathbf{C}^{(1)}_{22} &\mathbf{C}^{(3)}_{21}& \mathbf{C}^{(3)}_{22} \\
 \mathbf{C}^{(3) \rm T}_{11} & \mathbf{C}^{(3) \rm T}_{21} & \mathbf{C}^{(2)}_{11} & 0 \\
 \mathbf{C}^{(3) \rm T}_{12} & \mathbf{C}^{(3) \rm T}_{22} & 0 & \mathbf{C}^{(2)}_{22}
 \end{array}
 \right] ,\\
 \mathbf{S} &=& \left[
 \begin{array}{cccc}
  \mathbf{S}^{(1)}_{11} &0& 0 & 0 \\
 0 & \mathbf{S}^{(1)}_{22} & 0 & 0 \\
 0 & 0 & \mathbf{S}^{(2)}_{11} & 0 \\
 0 & 0 & 0 & \mathbf{S}^{(2)}_{22}
 \end{array}
 \right] ,
 \end{eqnarray}
where the matrices $\mathbf{C}^{(1,2,3)}_{ij}$ and $\mathbf{S}^{(1,2)}_{ij}$ are $N \times N $ matrices. Their explicit expressions are given in Appendix \ref{app:explicit_naive}.

In the KBBF technique, a special choice of basis functions is made: the basis for the large and small components are related by a transformation guaranteeing that the non-relativistic equation is recovered in the limit $c \rightarrow \infty$. This is achieved by considering the relation between the small and large component given in (\ref{eq:sm_comp}). Substituting (\ref{eq:basis_2}) in the latter, we get
\begin{eqnarray}
 \chi(\xi,\eta) = \frac{\hat{R}}{E+mc^{2} -V(\xi,\eta)} \left(
\begin{array}{c}
 \sum_{n=1}^{N} a_{n}^{(1)}B^{(1)}_{n}(\xi,\eta) \\
\sum_{n=1}^{N} a_{n}^{(2)}B^{(2)}_{n}(\xi,\eta)
\end{array}
\right). 
\end{eqnarray}
Note here that substituting this equation in the Rayleigh-Ritz coefficient allows recovering the min-max method and its discretization scheme described in the last section. Then, the eigensolutions have to be calculated by using an iteration procedure because the functional has an intrinsic dependence on the eigenenergy. This is circumvented by neglecting the space dependence of the potential over the support of each basis function and redefining the basis expansion coefficients as
\begin{eqnarray}
 c_{n}^{(1,2)} &=&  \frac{1}{E+mc^{2} -V_{n}} a_{n}^{(1,2)} ,\\
 & \approx & \frac{1}{E+mc^{2} -V(\xi,\eta)} a_{n}^{(1,2)} ,
\end{eqnarray}
where $V_{n}$ is a constant coefficient representing the contribution of the potential on the support of the basis function $n$.
Thus, we obtain
\begin{eqnarray}
\label{eq:basis_kbbf}
 \chi(\xi,\eta) = \frac{\hat{R}}{2mc^2} \left(
\begin{array}{c}
 \sum_{n=1}^{N} c_{n}^{(1)}B^{(1)}_{n}(\xi,\eta) \\
\sum_{n=1}^{N} c_{n}^{(2)}B^{(2)}_{n}(\xi,\eta)
\end{array}
\right).
\end{eqnarray}
where the factor $1/2mc^2$ was included for numerical convenience (it also allows to recover the non-relativistic limit exactly if the energy is shifted by $mc^{2}$). In some sense, the eigenenergy and space dependence of the prefactor $1/(E+mc^{2} -V)$ is encoded in the coefficients $c_{n}^{(1,2)}$. No iteration procedure is required as in the min-max method but the eigenvalue problem is larger. Also, the relation between the small and large components is only approximate because the constants $c_{n}^{(1,2)}$ have neither energy nor space dependence. However, we expect this relation to converge toward the exact one as the number of basis functions is increased and their support decreases. In that case, neglecting the spatial variation of the potential becomes a better approximation. In the limit $N \rightarrow \infty$, we have
\begin{eqnarray}
 \chi(\xi,\eta) = \frac{\hat{R}}{2mc^2} f(\xi,\eta) ,
\end{eqnarray}
where $f(\xi,\eta)$ is a bi-spinor. This implies that in the KBBF, the extremization of the Rayleigh quotient is performed on $f(\xi,\eta)$ rather than on $\chi(\xi,\eta)$ as in the ``naive'' Rayleigh-Ritz method. The former is consistent with the min-max principle exposed previously and the stationary point (or Euler-Lagrange equation) is given by
\begin{eqnarray}
\hat{R}\frac{E+mc^{2} -V}{2mc^2}\hat{R}f  =   \hat{R}^{2} \phi .
\end{eqnarray}
This equation can also be obtained from a unitary transformation of the Dirac equation \cite{QUA:QUA560250112}. Therefore, in the continuous limit, the exact solution is recovered from the min-max principle, which establishes that the two approaches are consistent with each other in that limit.

Explicitly, the basis function expansion is given in prolate spheroidal coordinates by (here, we dropped the basis function argument for simplicity)
\begin{eqnarray}
 \label{eq:b_KBBF_1}
 \phi_{1,2}(\xi,\eta) &=& \sum_{n=1}^{N} a_{n}^{(1,2)}B^{(1,2)}_{n} ,\\
\label{eq:b_KBBF_2}
\chi_{1}(\xi,\eta) &=& \frac{i}{2mc}\sum_{n=1}^{N} \biggl\{ c_{n}^{(2)}\left[ - \partial_{r} - \frac{\mu_{2}}{r}  \right]  B^{(2)}_{n} -   c_{n}^{(1)}\partial_{z} B^{(1)}_{n} \biggr\} ,\\
\label{eq:b_KBBF_3}
\chi_{2}(\xi,\eta) &=& \frac{i}{2mc} \sum_{n=1}^{N} \biggl\{ c_{n}^{(1)} \left[ - \partial_{r} + \frac{\mu_{1}}{r}  \right]  B^{(1)}_{n} +  c_{n}^{(2)} \partial_{z}B^{(2)}_{n} \biggr\} ,
\end{eqnarray}
where $\partial_{r}$ and $\partial_{z}$ are given in (\ref{eq:deri_r}) and (\ref{eq:deri_z}) respectively, while $r = R \left[ (\xi^{2} - 1)(1-\eta^{2}) \right]^{\frac{1}{2}}$. These last expression were obtained by expressing the operator $\hat{R}$ explicitly in prolate spheroidal coordinates.

Substituting (\ref{eq:b_KBBF_1}) to (\ref{eq:b_KBBF_3}) in (\ref{eq:ray_ritz}), we also obtain a generalized eigenvalue problem in the form of (\ref{eq:gen_eig}).
%
%
In this case, the matrix structure is
\begin{eqnarray}
 \mathbf{C}  &=& \left[
\begin{array}{cccc}
 \mathbf{C}^{(1)}_{11} &0&\mathbf{C}^{(3)}_{11}&\mathbf{C}^{(3)}_{12} \\
0 & \mathbf{C}^{(1)}_{22} &\mathbf{C}^{(3)}_{21}& \mathbf{C}^{(3)}_{22} \\
\mathbf{C}^{(3) \rm T}_{11} & \mathbf{C}^{(3) \rm T}_{21} & \mathbf{C}^{(2)}_{11} & \mathbf{C}^{(2)}_{12} \\
\mathbf{C}^{(3) \rm T}_{12} & \mathbf{C}^{(3) \rm T}_{22} & \mathbf{C}^{(2) \rm T}_{12} & \mathbf{C}^{(2)}_{22}
\end{array}
\right] ,\\
\mathbf{S} &=& \left[
\begin{array}{cccc}
 \mathbf{S}^{(1)}_{11} &0& 0 & 0 \\
0 & \mathbf{S}^{(1)}_{22} & 0 & 0 \\
0 & 0 & \mathbf{S}^{(2)}_{11} & \mathbf{S}^{(2)}_{12} \\
0 & 0 & \mathbf{S}^{(2) \rm T}_{12} & \mathbf{S}^{(2)}_{22}
\end{array}
\right] ,
\end{eqnarray}
and explicit expressions can be found in Appendix \ref{app:explicit_RRKBBF}. The matrix $\mathbf{C}$ and $\mathbf{S}$ are $4N\times 4N$ matrices while the other components ($\mathbf{C}^{(1,2,3)}_{ij}$ and $\mathbf{S}^{(1,2)}_{ij}$) are $N \times N$ matrices. 

The analytical analysis of convergence of this method and the proof that it does not have spurious solutions is clearly a non-trivial matter but was discussed in \cite{PhysRevA.62.022508} for the L-spinors basis functions. In our case, these properties will be verified empirically by looking at the numerical results while a careful analysis of the method is currently under investigation.

\subsection{Basis functions}

Throughout this work, B-spline basis functions are used (a description of these functions can be found in \cite{0034-4885-64-12-205}). This choice is favored over other techniques because it can be easily implemented and because B-splines have compact support, leading to sparse matrix structures. This allows using powerful numerical routines for the calculation of eigenvalues. More important is the fact that B-splines are linearly independent and form a complete basis, which is a necessary condition for the convergence of the Rayleigh-Ritz method for both eigenenergies and eigenfunctions (\cite{PhysRevA.62.022508} and references therein). It is also an important requirement to avoid errors of order $1/c^{4}$ in the Rayleigh-Ritz bounds which may induce spurious states in certain circumstances \cite{10.1063/1.447865}.

B-splines basis functions have been studied extensively for solving the time-independent Dirac equation because of these important properties. However, most of these studies considered atoms or atomic-like systems \cite{PhysRevA.37.307,PhysRevLett.93.130405,Dolbeault2003,JPSJ.75.114301}, although recently, they were used for diatomic molecules \cite{Dolbeault2003,0953-4075-43-23-235207}.  

\begin{figure}[h]
	\centering
	\includegraphics[width=0.70\textwidth]{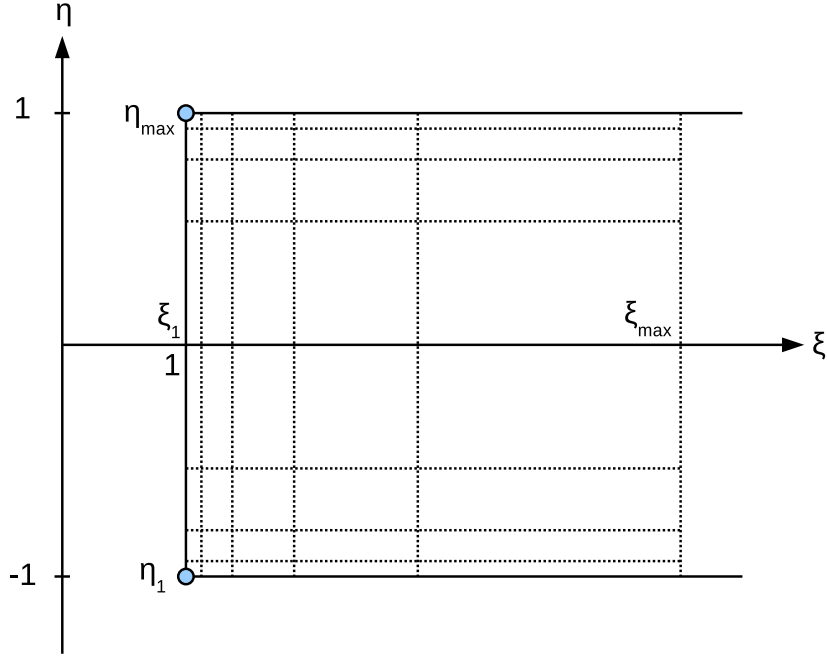}
	\caption{Discretization of the domain. A circle represents a point where there is a Coulomb singularity. The domain under consideration is discretized into a certain number of elements while each element is subdivided into smaller regions for the space integration. Note that we are using an exponential size distribution with smaller elements close to the Coulomb singularities. There is a knot point at every intersection of two dotted lines.}
	\label{fig:Bspline_dis}
\end{figure}

B-splines are fully determined by their order $k_{\xi,\eta}$ and knot vector using the iterative formula \cite{0034-4885-64-12-205,nla.cat-vn991654}
\begin{eqnarray}
b_{i}^{k}(x) = \frac{x-t_{i}}{t_{i+k-1} - t_{i}}b^{k-1}_{i}(x) + \frac{t_{i+k} -x}{t_{i+k}-t_{i+1}} b^{k-1}_{i+1}(x) ,
\end{eqnarray}
and initial conditions
\begin{eqnarray}
 b_{i}^{1}(x) = 1 \;\;\mbox{for} \; \;t_{i} \leq x < t_{i+1} \;\; \mbox{and} \;\; b_{i}^{1} = 0 \;\; \mbox{otherwise},
\end{eqnarray}
where $t_{i}$'s are knot coordinates. The number of knots at a given coordinate determines the continuity condition at that point. Therefore, the number of knots should be maximal at singular points (at the Coulomb singularity position for instance) to allow for a discontinuous behavior of the wave function. Throughout this work, the knot vectors are given by the sequences 
\begin{eqnarray}
1 = \xi_{1} = ... = \xi_{k_{\xi}} < \xi_{k_{\xi}+1} < ...< \xi_{n_{\xi}+1} = ... = \xi_{n_{\xi}+k_{\xi}} = \xi_{\rm max} ,\\
-1 = \eta_{1} = ... = \eta_{k_{\eta}} < \eta_{k_{\eta}+1} < ...< \eta_{n_{\eta}+1} = ... = \eta_{n_{\eta}+k_{\eta}} = 1 .
\end{eqnarray}
Here, $n_{\xi,\eta} $ are the number of spline functions in $\xi$ and $\eta$ coordinates, respectively. The knot coordinates can be chosen arbitrarily in the domain under consideration. However, to improve accuracy, an exponential sequence with smaller intervals close to the singularities is used in this study. The knot sequences and domain structure for diatomic molecules are depicted in Figure \ref{fig:Bspline_dis}.

The basis function can then be written as the tensor product of B-spline functions as
\begin{eqnarray}
\label{eq:basis_func_def}
B^{(1,2)}_{n}(\xi,\eta) = G^{(1,2)}(\xi,\eta) b_{i}^{k_{\xi}}(\xi)b_{j}^{k_{\eta}}(\eta) ,
\end{eqnarray}
where $n = [i,j] \in \mathbb{Z}^{2}$, $i \in [1,n_{\xi}]$ and $j \in [1,n_{\eta}]$. 
The overall factor is defined as \cite{Kullie2004215,Kullie1999307}
%
%
%
\begin{eqnarray}
G^{(1,2)}(\xi,\eta) = [(\xi^{2}-1)(1-\eta^{2})]^{\frac{\mu_{1,2}}{2}} ,
\end{eqnarray}
where
\begin{eqnarray}
 \mu_{1} = j_{z}-\frac{1}{2} \;\; \mbox{and} \;\; \mu_{2} = j_{z}+\frac{1}{2}.
\end{eqnarray}
This factor accounts for the angular momentum dependence (remember that $j_{z}$ is the angular momentum projection on the z-axis). Moreover, it allows having well-defined integrals in the functionals, allowing a better convergence of the method. 


\subsection{Details of the calculation}

The construction of the matrices appearing in the two numerical methods involves several integrals extending over the whole domain. However, because B-splines are compact, the integration domains are reduced to the support of each basis function, which are regions having $k_{\xi} \times k_{\eta}$ elements or less. The integrals are evaluated numerically using the Gauss-Legendre quadrature rule.

The boundary conditions are chosen as $\phi(\xi_{\max},\eta) = 0$ and $\chi(\xi_{\max},\eta) = 0$. Using B-splines, this can be implemented easily by setting $b_{n_{\xi}}(\xi) = 0$ and by considering only $n_{\xi}-1$ B-spline functions in $\xi$ coordinates. The other boundaries are free. Strictly speaking, these boundary conditions may lead to some numerical problems because of the occurrence of the Klein paradox \cite{PhysRevA.37.307}: they correspond to the confining of the electron in a box surrounded by an infinite potential barrier at $\xi = \xi_{\rm max}$. However, similar conditions have been used successfully in \cite{PhysRevA.48.2700,Kullie1999307,Kullie2004215} to obtain very accurate results. Also, it was argued in \cite{Dolbeault2003,PhysRevLett.93.130405} that using these zero boundary conditions have negligible effects on the solution if the domain is large enough.

The boundary conditions on the nuclei are more subtle. It is argued in \cite{0953-4075-42-5-055002} that B-spline basis expansion are plagued with non-physical solutions related to a wrong treatment of these boundary conditions. A remedy to this problem is also proposed but it is not clear to us how this technique can be applied to the diatomic molecule case. However, it is verified \textit{a posteriori} by looking at the spectrum that our boundary conditions does not induce spurious states.


The code performing the calculation is parallelized by using the domain decomposition strategy described in \cite{fillion2011aa}. For better performance, the ScaLAPACK library is utilized to solve the eigenvalue and generalized eigenvalue problems. In the Rayleigh-Ritz method, the latter yields the whole energy spectrum and eigenfunction in one calculation. For the min-max method, only one eigenenergy can be calculated at a time because each evaluation necessitates a solution of $\Lambda_{k}(E)=0$. The latter is solved by a root-finding algorithm based on Brent's method \cite{brent1973}.

\section{Results and discussion}
\label{sec:res}

In this section, the results obtained for both numerical methods are presented. First, the convergence of the method is analyzed. Then, the spectra of diatomic molecules are presented and finally, the absence of spurious states is discussed.

\subsection{Convergence of the method}

In this section, we are investigating the convergence of our numerical methods.  More specifically, we study and calculate the ground state of dithorium ($\mbox{Th}_{2}^{179+}$ which has $Z_{1,2}=90$) and dihydrogen ($\mbox{H}_{2}^{+}$ which has $Z_{1,2}=1$)). The semi inter atomic distance is set to $R = \frac{1}{90} \approx 0.011111$ a.u. for dithorium and to $R \approx 1.000$ a.u. for dihydrogen while the angular momentum is taken as $j_{z} = 1/2$. The results for the calculation of the ground state binding energy using B-splines of order 7 and different mesh sizes are shown in Table \ref{table:res_conv_H2} and \ref{table:res_conv_Th179} for $\mbox{H}_{2}^{+}$ and $\mbox{Th}_{2}^{179+}$, respectively.

The results presented in this table show the convergence of the method as the number of elements is increased. The results obtained are very accurate, although there is a small relative difference $d_{r}$ between our results and the results presented in \cite{Kullie2004215}. For $\mbox{H}_{2}^{+}$, we obtain $d_{r} \approx 2.2 \times 10^{-8}$\% and $d_{r} \approx 1.9 \times 10^{-8}$\% for the min-max and Rayleigh-Ritz methods, respectively, while for $\mbox{Th}_{2}^{179+}$, we have $ d_{r} \approx 6.1 \times 10^{-4}$\% and $ d_{r} \approx 0.97 \times 10^{-4}$\% for the min-max and Rayleigh-Ritz methods, respectively. These differences can be explained by a different choice of boundary conditions, different element formulation and different treatment of the Coulomb singularity.  

In all cases, the convergence is from above suggesting that both methods are consistent with the min-max principle. The convergence in the case of dithorium however is much slower than for dihydrogen. One possible reason explaining this discrepancy is the behavior of B-splines close to the Coulomb singularities. In that region, the wave function should behave like $\psi \sim r_{1,2}^{-1+\gamma_{1,2}}$ (obtained in atomic calculations) where
\begin{eqnarray}
\gamma_{1,2} = \sqrt{\left(|j_{z}| + \frac{1}{2}\right)^{2} - \alpha^{2} Z_{1,2}^{2}} .
\end{eqnarray}
and $r_{1,2}$ are the distances from nuclei 1 and 2. In ground state calculations, we have $j_{z} = 1/2$ and thus, $0 < \gamma_{1,2} < 1$ for $Z_{1,2}<137$. Therefore, the wave function has a non-integer power-law behavior close to the singularity. The B-spline basis functions, being polynomial with integer powers, are unable to reproduce exactly this feature. Moreover, we have that
\begin{eqnarray}
 \gamma_{\rm H} \approx 0.999947 \;\; \mbox{and} \;\; \psi \sim r_{1,2}^{-0.000053} ,\\
\gamma_{\rm Th} \approx 0.568664 \;\; \mbox{and} \;\; \psi \sim r_{1,2}^{-0.431336} ,
\end{eqnarray}
where $\gamma_{\rm H,Th}$ are the gamma associated with a hydrogen and thorium atom. It is clear from this that the behavior of the wave function is much closer to a power law for dihydrogen and therefore, is better reproduced by the B-splines and also, has a faster convergence. 

One possible cure to this is to use another prefactor in the basis function that captures the correct behavior. For instance, it was proposed to multiply the basis functions in (\ref{eq:basis_func_def}) by \cite{PhysRevA.48.2700,Kullie1999307,Kullie2004215}
\begin{eqnarray}
G'(\xi,\eta) = r_{1}^{-1+\gamma_{1}}  r_{2}^{-1+\gamma_{2}} ,
\end{eqnarray}
with
\begin{eqnarray}
 r_{1} = (\xi+\eta)R, \; r_{2} = (\xi-\eta)R .
\end{eqnarray}
The main issue with this technique is that derivatives in the functionals become singular at the nuclei position. To cope with this, a singular coordinate transformation can be performed that allows transforming the singular non-integer behavior near the nuclei to a polynomial approximation \cite{PhysRevA.48.2700,Kullie1999307}. The latter can then be fitted more accurately with a polynomial basis function. We do not implement this technique here as the goal of this paper is not to achieve the most accurate value of bound state energies. However, it can be done in principle and could improve the convergence of the numerical method.

\begin{table}[h]
\caption{Results of the numerical computation for the ground state of $\mbox{H}_{2}^{+}$ for different mesh sizes and B-spline of order 7. Here, $N_{\xi,\eta}$ are the number of elements in each coordinates while $N^{*}$ is the total number of basis functions utilized. The maximum coordinate was fixed to $\xi_{\rm max}=30$ a.u. and the angular momentum to $j_{z} = 1/2$. The calculations are to be compared with the results from \cite{Kullie2004215} where the authors obtained E$_{\mathrm{H}_{2}^{+}}$ = -1.10264158103 a.u..}
\centering
\begin{tabular}{lllcc}
 \hline \hline
$N_{\xi}$ & $N_{\eta}$ & $N^{*}$  &  \multicolumn{2}{c}{E$_{\mathrm{H}_{2}^{+}}$ (a.u.)} \\
	  &	       &          & Min-max & RR \\
 \hline  
8 & 8 & 182 & -1.102590816884 & -1.102590816895 \\
10 & 10 & 240 & -1.102638533873 & -1.102638533934 \\
12 & 12 & 306 & -1.102641366239 & -1.102641366228 \\
14 & 14 & 380 & -1.102641554428 & -1.102641554501 \\
16 & 16 & 462 & -1.102641577089 & -1.102641577085 \\
18 & 18 & 552 & -1.102641580210 & -1.102641580229 \\
20 & 20 & 650 & -1.102641580782 & -1.102641580825 \\
 \hline \hline 
\end{tabular} 
\label{table:res_conv_H2}
\end{table}

\begin{table}[h]
\caption{Results of the numerical computation for the ground state of $\mbox{Th}_{2}^{179+}$ for different mesh sizes and B-spline of order 7. Here, $N_{\xi,\eta}$ are the number of elements in each coordinates while $N^{*}$ is the total number of basis functions utilized. The maximum coordinate was fixed to $\xi_{\rm max}=15$ a.u. and the angular momentum to $j_{z} = 1/2$. The calculations are to be compared with the results from \cite{Kullie2004215} and \cite{0953-4075-43-23-235207} where the authors obtained E$_{\mathrm{Th}_{179}^{+}}$ = -9504.756746922 a.u. and E$_{\mathrm{Th}_{179}^{+}}$ = -9504.752 a.u..}
\centering
\begin{tabular}{lllcc}
 \hline \hline
$N_{\xi}$ & $N_{\eta}$ & $N^{*}$  &  \multicolumn{2}{c}{E$_{\mathrm{Th}_{2}^{179+}}$  (a.u.)} \\
	  &	       &          & Min-max & RR \\
 \hline  
8  & 8  & 182 & -9503.998584802 & -9504.592867005\\
10 & 10 & 240 & -9504.333585765 & -9504.687658554\\
12 & 12 & 306 & -9504.466070634 & -9504.711111628\\
14 & 14 & 380 & -9504.539502492 & -9504.722791962\\
16 & 16 & 462 & -9504.586247153 & -9504.730034585\\
18 & 18 & 552 & -9504.618392312 & -9504.735005730\\
20 & 20 & 650 & -9504.641636959 & -9504.738611929\\
24 & 24 & 870 & -9504.672557123 & -9504.743429586\\
30 & 30 & 1260& -9504.698874401 & -9504.747552293\\
 \hline \hline 
\end{tabular} 
\label{table:res_conv_Th179}
\end{table}

\subsection{Spectra of diatomic molecules}

In this section, the spectra of dihydrogen and dithorium are presented. They are shown in Tables \ref{table:res_spec_H2} and \ref{table:res_spec_Th2} for $j_{z}=1/2$. The spectra are calculated using a mesh of 14$\times$14 elements and 17$\times$17 elements for the min-max method for dihydrogen and dithorium, respectively, while a mesh of 30$\times$30 elements is utilized in the Rayleigh-Ritz method. The other parameters are set to the same values as in the last section where the convergence of the ground state was discussed. The binding energy values in the mass gap $(-mc^{2},mc^{2})$, corresponding to bound states, are shifted by $mc^{2}$ to have a comparison with non-relativistic results. The values in the continua however are not shifted and calculated with the Rayleigh-Ritz method only. The results of the dithorium spectrum can be compared to the ones in \cite{0953-4075-37-4-016}. Both are generally in good agreement, although a small discrepancy can be seen for the higher excited states.

In the Rayleigh-Ritz method, the $n_{\rm binding}$ bound state energies shown in Tables \ref{table:res_spec_H2} and \ref{table:res_spec_Th2} correspond to the $2N+1$ to $2N+1+n_{\rm binding}$ eigenvalues of the matrix $\mathbf{C}$ (once the eigenvalues are ordered in increasing order). The other eigenvalues can be associated to the ``discretized'' negative (the first to the $2N$'th eigenvalues) and positive (the $2N+2+n_{\rm binding}$'th to the $4N$'th eigenvalues) energy continua. For the min-max method, the bound state energies shown corresponds to the solution of $\Lambda_{k}(E)=0$ for the $n_{\rm binding}$ lowest energy eigenvalues. For the diatomic molecules considered, the spectra calculated with both methods are in very good agreement. The small discrepancy remaining is mostly due to the use of different mesh sizes.  

The convergence of the excited states is very similar to one of the ground state: all values are approached from above and the order of convergence is close to the one of the ground state. The same is true for the states in the positive energy continuum, that is for $E \geq mc^{2}$. For the negative energy states, the convergence occurs from below, but otherwise, follows the same trends as the other cases. The energy values in the continua (especially their smallest and largest eigenvalues) depend on the size of the domain. In the dithorium calculation, the domain was smaller which yielded less accurate value in the continua (not shown in the table) but better accuracy of the bound states. In all cases, the eigenvalues of the positive and negative energy continua accumulate at the points $mc^{2}$ and $-mc^{2}$, respectively.

\begin{table}[h]
\caption{Results of the numerical computations for the spectrum of $\mbox{H}_{2}^{+}$ for a mesh size of 30$\times$ 30 and B-spline of order 7. The states of the positive and negative continua are computed with the Rayleigh-Ritz method and only the first 25 states are shown.}
\centering
\begin{tabular}{lll|lll}
 \hline \hline
Bound &  \multicolumn{2}{c}{Binding energy  (a.u.)} & & Negative & Positive\\
states& Min-max & \multicolumn{2}{l}{RR} &continuum (a.u.) & continuum (a.u.)\\
 \hline 
1 & -1.1026413662& -1.1026415808 & 1& -18778.95240& 18778.86549\\
2 & -0.6675525594& -0.6675527718 & 2& -18778.95792& 18778.86561\\
3 & -0.4287795568& -0.4287811584 & 3& -18778.96471& 18778.86562\\
4 & -0.3608697621& -0.3608710695 & 4& -18778.97284& 18778.86741\\
5 & -0.2554175614& -0.2554197033 & 5& -18778.98233& 18778.86746\\
6 & -0.2357807609& -0.2357812681 & 6& -18778.98475& 18778.86808\\
7 & -0.2267021482& -0.2267030696 & 7& -18778.99077& 18778.86917\\
8 & -0.2008621355& -0.2008689095 & 8& -18778.99320& 18778.88272\\
9 &  -0.1776816232& -0.1776839788 &9 & -18778.99684& 18778.88275\\
10 & -0.1373089205& -0.1373147686 &10 & -18778.99979& 18778.88617\\
11 & -0.1307908409& -0.1307928214 & 11& -18779.00397& 18778.88627\\
12 & -0.1267066133& -0.1267100818 & 12& -18779.00544& 18778.88659\\
13 & -0.1266438351& -0.1266441499 & 13& -18779.00643& 18778.89067\\
14 & -0.1261986510& -0.1261992440 & 14& -18779.01259& 18778.89068\\
15 & -0.1158897902& -0.1159009024 & 15& -18779.01542& 18778.89977\\
16 & -0.1053558675& -0.1053611251 & 16& -18779.01903& 18778.89987\\
17 & -0.0852450505& -0.0852548082 & 17& -18779.02284& 18778.90406\\
18 & -0.0823477309& -0.0823523149 & 18& -18779.02659& 18778.90575\\
19 & -0.0804553251& -0.0804564514 & 19& -18779.03400& 18778.90766\\
20 & -0.0802631100& -0.0802631662 & 20& -18779.03472& 18778.91436\\
21 & -0.0802102415& -0.0802110614 & 21& -18779.03980& 18778.91438\\
22 & -0.0802047983& -0.0802048234 & 22& -18779.04047& 18778.91544\\
23 & -0.0800201297& -0.0800252078 & 23& -18779.04625& 18778.91591\\
24 & -0.0730502242& -0.0730676123 & 24& -18779.04822& 18778.91599\\
25 & -0.0649840057& -0.0649993038 & 25& -18779.05030& 18778.91605\\
 \hline \hline 
\end{tabular} 
\label{table:res_spec_H2}
\end{table}

%
%

\begin{table}[h]
\caption{Results of the numerical computation for the spectrum of $\mbox{Th}_{2}^{179+}$. The mesh size is indicated on the second line. The B-splines are of order 7.}
\centering
\begin{tabular}{lllll}
 \hline \hline
States &  \multicolumn{2}{l}{Naive RR} &  RR & Min-max\\
	& $14\times14$ & $30 \times 30$ & $30 \times 30$ & $16 \times 16$ \\
 \hline 
1 & -9504.6525442 & -9504.7243225 & -9504.7475523 & -9504.5862992\\ 
2 & -6815.3652913 & -6815.4657298 & -6815.5599111 & -6815.3230307\\ 
3 & -4127.8799531 & -4127.8877478 & -4128.1451137 & -4127.8197047\\ 
4 & -3374.4958326 & -3374.5117016 & -3374.5143753 & -3374.4569981\\ 
5 & -2564.1326367 & -2564.1559253 & -2564.1719708 & -2564.0744037\\ 
6 & -2455.9453341 & -2455.9537953 & -2455.9600280 & -2455.8837393\\ 
7 & -2010.6579407 & -2010.6535604 & -2010.4321103 & -2010.4241948\\ 
8 & -1918.5275474 & -1918.4056980 & -1915.7178408 & -1915.6761267\\ 
9 & -1649.5111100 & -1649.2929148 & -1643.9543595 &  -1643.9320665\\ 
10 & -1349.5529034 & -1344.0855870 & -1313.8071916 & -1313.7606899\\ 
11 & -1339.1123032 & -1333.5368147 & -1303.6850950 & -1303.6580541\\ 
spurious & -1218.2113620 & -1204.6990945 &  \\ 
12 & -1169.3956263 & -1159.1761393 & -1089.6415827 & -1089.6356220\\ 
13 & -1138.5709512 & -1131.0151665 & -1084.3699127 & -1084.3519981\\ 
14 & -1046.2053120 & -1045.4764538 & -1028.1920826 & -1028.1912423\\ 
15 & -1018.4013912 & -984.5252901 & -969.6816867 & -969.64172165\\ 
 \hline \hline 
\end{tabular} 
\label{table:res_spec_Th2}
\end{table}

\subsection{Spurious states}

The results for the spectra of diatomic molecules presented in the last section showed a spurious state in the dithorium spectrum calculated with the naive Rayleigh-Ritz method while the other methods did not. Spurious states usually appear as eigenstates with an energy in the interval $-mc^{2} < E < E_{\rm ground}$ because of their highly oscillatory behavior. This was not observed in the numerical results. Moreover, it was proven mathematically that the min-max method is free from these numerical artifacts \cite{Dolbeault2003}. The spectra predicted by the min-max and the Rayleigh-Ritz methods coincides (up to numerical errors), implying that our version of the Rayleigh-Ritz method using kinematically balanced function is also free from these unphysical states.

These last arguments are mainly qualitative. A more convincing approach proceeds by computing the spectrum for an atom (by setting $Z_{2}=0$) and by comparing to the well-known analytical formula for the atomic binding energy given by \cite{Itzykson:1980rh}
\begin{eqnarray}
 E_{nj} = \frac{mc^{2}}{\sqrt{1+\frac{Z^{2}\alpha^{2}}{(n-\delta_{j})^{2}}}} - mc^{2} ,
\end{eqnarray}
where $n$ is the principal quantum number, $j$ is the angular momentum and
\begin{eqnarray}
 \delta_{j} = j + \frac{1}{2} - \sqrt{\left(j+\frac{1}{2} \right) - Z^{2}\alpha^{2}}.
\end{eqnarray}
Of course, the numerical methods are not optimized for atomic calculations, but these results, albeit not very accurate, allow showing that no spurious states appear. The results for the spectrum of Th$^{89+}$ are shown in Table \ref{table:res_spec_Th}. For all eigenenergies considered, there is always a one-to-one correspondence between the analytical and the Rayleigh-Ritz results, in contradistinction with the spectrum obtained from the naive Rayleigh-Ritz method which exhibits spurious states.  

\newpage

\begin{longtable}{llll}
\caption{Results for the spectrum of $\mbox{Th}^{89+}$. The mesh size is 30$\times$30 and the B-splines are of order 7. The states are denoted in spectroscopic notation.}
 \\ \hline 
 \hline 
States &  Analytical &  RR & Naive RR\\
 & (a.u.) & (a.u.) & (a.u.) \\
 \hline 
1s$_{\frac{1}{2}}$ & -4617.757542 & -4615.302929 & -4636.678774 \\ 
2s$_{\frac{1}{2}}$ & -1192.289212 & -1192.108524 & -1201.301342 \\ 
2p$_{\frac{1}{2}}$ & -1192.289212 & -1191.771020 & -1192.051000 \\
2p$_{\frac{1}{2}}$ & -1041.374505 & -1041.374468 & -1041.383185 \\ 
3s$_{\frac{1}{2}}$ & -512.199990 & -512.140001 & -526.114688 \\ 
3p$_{\frac{1}{2}}$ & -512.199990 & -512.038910 & -512.120856 \\ 
3p$_{\frac{1}{2}}$ & -467.182486 & -467.182452 & -467.295571 \\ 
3d$_{\frac{1}{2}}$ & -467.182486 & -467.182410 & -467.182518 \\ 
spurious &  &  & -462.487121 \\ 
3d$_{\frac{3}{2}}$ & -455.524906 & -455.524869 & -455.524983 \\ 
spurious &  &  & -341.014173 \\ 
4s$_{\frac{1}{2}}$ & -280.938972 & -280.913251 & -281.417745 \\ 
4p$_{\frac{1}{2}}$ & -280.938972 & -280.871281 & -280.904780 \\ 
4p$_{\frac{1}{2}}$ & -262.173744 & -262.173642 & -262.173904 \\ 
4d$_{\frac{1}{2}}$ & -262.173744 & -262.173503 & -262.173811 \\ 
4d$_{\frac{3}{2}}$ & -257.210164 & -257.210092 & -257.210222 \\ 
4f$_{\frac{3}{2}}$ & -257.210164 & -257.209987 & -257.209640 \\ 
4f$_{\frac{5}{2}}$ & -254.854358 & -254.854272 & -254.854328 \\ 
spurious &  &  & -195.443773 \\ 
5s$_{\frac{1}{2}}$ & -176.667335 & -176.653978 & -176.650586 \\ 
5p$_{\frac{1}{2}}$ & -176.667335 & -176.633024 & -175.916124 \\ 
5p$_{\frac{1}{2}}$ & -167.174184 & -167.173989 & -167.177456 \\ 
5d$_{\frac{1}{2}}$ & -167.174184 & -167.173578 & -167.174285 \\ 
5d$_{\frac{3}{2}}$ & -164.630108 & -164.629880 & -164.876895 \\ 
5f$_{\frac{3}{2}}$ & -164.630108 & -164.629582 & -164.630142 \\ 
5f$_{\frac{5}{2}}$ & -163.417654 & -163.417526 & -164.438475 \\ 
5g$_{\frac{5}{2}}$ & -163.417654 & -163.417279 & -163.417704 \\ 
spurious &  &  & -163.375298 \\ 
5g$_{\frac{7}{2}}$ & -162.704858 & -162.704687 & -162.694189 \\ 
6s$_{\frac{1}{2}}$ & -121.138528 & -121.130514 & -121.774091 \\ 
6p$_{\frac{1}{2}}$ & -121.138528 & -121.118841 & -121.128677 \\ 
6p$_{\frac{1}{2}}$ & -115.699215 & -115.698928 & -115.700224 \\ 
6d$_{\frac{1}{2}}$ & -115.699215 & -115.698194 & -115.699014 \\ 
6d$_{\frac{3}{2}}$ & -114.228643 & -114.228211 & -114.228851 \\ 
6f$_{\frac{3}{2}}$ & -114.228643 & -114.227668 & -114.225881 \\ 
6f$_{\frac{5}{2}}$ & -113.525840 & -113.525445 & -113.526110 \\ 
6g$_{\frac{5}{2}}$ & -113.525840 & -113.524951 & -113.523149 \\ 
6g$_{\frac{7}{2}}$ & -113.112069 & -113.111850 & -113.112958 \\ 
6h$_{\frac{7}{2}}$ & -113.112069 & -113.111368 & -113.110337 \\ 
spurious &  &  & -112.924774 \\
6h$_{\frac{9}{2}}$ & -112.839015 & -112.838682 & -112.839037 \\  
spurious &  &  & -104.197448 \\ 
 \hline \hline 
\label{table:res_spec_Th}
\end{longtable}

\section{Conclusion}
\label{sec:conclusion}

In this work, we presented two numerical methods to solve the single particle time-independent Dirac equation. The first one was based on a min-max variational principle while the second one used a combination of the Rayleigh-Ritz method and kinematically balanced basis functions. For comparison purposes, we also included a description of the naive Rayleigh-Ritz method. All were based on a B-spline basis function discretization which allowed obtaining a high accuracy and a sparse matrix structure in the discretized equations. We applied these methods to the computation of the two-centre Coulomb problem ground state energy and spectrum. Because of its axial symmetry and simple structure, it was convenient to use prolate spheroidal coordinates.  These techniques were used specifically to compute the spectra of the molecule $\mbox{H}_{2}^{+}$ and the quasi-molecule $\mbox{Th}_{2}^{179+}$. A comparison with results in the literature for the ground state demonstrated that our methods yield very accurate and convergent results, especially for dihydrogen. More importantly, no spurious states were reported in these numerical schemes and thus, they both could be used to evaluate radiative QED corrections for diatomic molecules which necessitate sums over intermediate states. This conclusion was reached by comparing the calculated spectra of dithorium and thorium to results obtained from the naive Rayleigh-Ritz method.

The two methods have strengths and weaknesses. In terms of computation time, the Rayleigh-Ritz methods were much faster, especially for the computation of the whole spectrum. This happens because in the min-max method, the solution of the eigenvalue equation $\Lambda_{k}(E)=0$ necessitates many iterations (typically between 20 and 30) to obtain a decent accuracy and each iteration requires a solution of the eigenvalue problem. This could be improved somewhat by using an iterative eigensolver optimized for the computation of few eigenvalues. Therefore, if one is only interested in the computation of the first few excited states while using a very large mesh, it may be advantageous to use the min-max method combined with a version of these iterative eigensolvers. In terms of accuracy, both methods yielded very similar results, although the convergence was slightly better for the Rayleigh-Ritz method in the dithorium case. This is in contradiction with the conclusion reached in \cite{0953-4075-38-16-008} where the min-max method showed much better accuracy. This discrepancy may be explained by a slightly different choice of basis functions (see (\ref{eq:basis_kbbf}) versus (6) of \cite{0953-4075-38-16-008}). Nevertheless, the main advantage of the Rayleigh-Ritz method is the fact that it gives the whole spectrum directly from the solution of the generalized eigenvalue problem. 

These methods can then be utilized in many applications. Among others, this can be used to investigate relativistic laser-matter interaction: the solution obtained from these methods can be used as an initial condition for the solution of the time-dependent Dirac equation. This will be the subject of future investigations.

\appendix

\section{Explicit expression for min-max method}
\label{app:explicit_minmax}

The explicit expression of matrices $\mathbf{A}_{11},\mathbf{A}_{22}$ and $\mathbf{A}_{12}$ is obtained by using the Dirac equation in cylindrical coordinates given in \cite{0305-4470-16-9-024}. Then, combining the ansatz in Eq. (\ref{eq:ansatz}) with the basis function expansion, we get
\begin{eqnarray}
\label{eq:explicit}
\left[\mathbf{A}_{11}\right]_{ij} & = &  \int d^{3}x \biggl\{ \biggl[ (\partial_{z}B^{(1)}_{i})(\partial_{z}B^{(1)}_{j}) 
+(\partial_{r}B^{(1)}_{i})(\partial_{r}B^{(1)}_{j}) \nonumber \\
&&\quad \quad \quad + \frac{\mu_{1}^{2}}{r^{2}} B^{(1)}_{i}B^{(1)}_{j}  - \frac{\mu_{1}}{r} B^{(1)}_{i}(\partial_{r}B^{(1)}_{j}) \nonumber \\
&&\quad \quad \quad - \frac{\mu_{1}}{r} (\partial_{r}B^{(1)}_{i})B^{(1)}_{j} \biggr] \frac{c^{2}}{E+mc^{2}-V} \nonumber \\
&&\quad \quad \quad + (V+mc^{2} -E)B^{(1)}_{i}B^{(1)}_{j} \biggr\} ,\\
\left[\mathbf{A}_{22}\right]_{ij} & = &  \int d^{3}x \biggl\{ \biggl[ (\partial_{z}B^{(2)}_{i})(\partial_{z}B^{(2)}_{j}) 
+(\partial_{r}B^{(2)}_{i})(\partial_{r}B^{(2)}_{j}) \nonumber \\
&&\quad \quad \quad + \frac{\mu_{2}^{2}}{r^{2}} B^{(2)}_{i}B^{(2)}_{j} + \frac{\mu_{2}}{r} B^{(2)}_{i}(\partial_{r}B^{(2)}_{j}) \nonumber \\
&&\quad \quad \quad + \frac{\mu_{2}}{r} (\partial_{r}B^{(2)}_{i})B^{(2)}_{j} \biggr] \frac{c^{2}}{E+mc^{2}-V} \nonumber \\
&&\quad \quad \quad +(V+mc^{2} -E) B^{(2)}_{i}B^{(2)}_{j} \biggr\} ,\\
\left[\mathbf{A}_{12}\right]_{ij} & = &  \int d^{3}x \biggl\{ \biggl[ (\partial_{z}B^{(1)}_{i})(\partial_{r}B^{(2)}_{j}) 
+ \frac{\mu_{1}}{r} B^{(1)}_{i}(\partial_{z}B^{(2)}_{j})  \nonumber \\ 
&&\quad \quad \quad  -(\partial_{r}B^{(1)}_{i})(\partial_{z}B^{(2)}_{j})
+ \frac{\mu_{2}}{r} (\partial_{z}B^{(1)}_{i})B^{(2)}_{j} \biggr] \nonumber \\
&& \quad \quad \quad \times \frac{c^{2}}{E+mc^{2}-V} \biggr\}.
\end{eqnarray}
The last expression can then be expressed in prolate spheroidal coordinates by using
\begin{eqnarray}
\label{eq:deri_r}
 \partial_{r} &=& \frac{\sqrt{(\xi^{2}-1)(1-\eta^{2})}}{R(\xi^{2}-\eta^{2})} \left[ \xi \partial_{\xi} - \eta \partial_{\eta} \right], \\
\label{eq:deri_z}
\partial_{z} &=& \frac{(\xi^{2}-1)}{R(\xi^{2}-\eta^{2})} \eta \partial_{\xi} + \frac{(1-\eta^{2})}{R(\xi^{2}-\eta^{2})} \xi \partial_{\eta},
\end{eqnarray}
and the integration measure is given by
\begin{eqnarray}
\label{eq:meas_pro_coord}
 d^{3}x = R^{3}(\xi^{2}-\eta^{2})d\xi d\eta d \theta .
\end{eqnarray}

 \section{Explicit expression for the naive Rayleigh-Ritz method}
 \label{app:explicit_naive}
 
 The explicit expression of matrices $\mathbf{C}^{(1)},\mathbf{C}^{(2)}$ and $\mathbf{C}^{(3)}$ is obtained by starting with the Dirac equation in cylindrical coordinates. By assuming that the basis functions are the same for the large and small components, we get
 \begin{eqnarray}
 \label{eq:explicit_nRR1}
 \bigl[\mathbf{C}^{(1)}_{11}\bigr]_{ij} & = &  \int d^{3}x \biggl\{ (V+mc^{2})B^{(1)}_{i}B^{(1)}_{j} \biggr\}, \\
 \bigl[\mathbf{C}^{(1)}_{22}\bigr]_{ij} & = &  \int d^{3}x \biggl\{ (V+mc^{2})B^{(2)}_{i}B^{(2)}_{j} \biggr\}, \\
 \bigl[\mathbf{C}^{(2)}_{11}\bigr]_{ij} & = &  \int d^{3}x \biggl\{ (V-mc^{2})B^{(1)}_{i}B^{(1)}_{j} \biggr\}, \\
 \bigl[\mathbf{C}^{(2)}_{22}\bigr]_{ij} & = &  \int d^{3}x \biggl\{ (V-mc^{2})B^{(2)}_{i}B^{(2)}_{j} \biggr\}, \\
 \bigl[\mathbf{C}^{(3)}_{11}\bigr]_{ij} & = &  \int d^{3}x \biggl\{ B^{(1)}_{i} \partial_{z} B^{(1)}_{j} \biggr\}, \\
 \bigl[\mathbf{C}^{(3)}_{22}\bigr]_{ij} & = &  -\int d^{3}x \biggl\{ B^{(2)}_{i} \partial_{z} B^{(2)}_{j} \biggr\}, \\
 \bigl[\mathbf{C}^{(3)}_{21}\bigr]_{ij} & = &  \int d^{3}x \biggl\{ B^{(2)}_{i} \biggl[ \partial_{r} - \frac{\mu_{1}}{r} \biggr] B^{(1)}_{j} \biggr\}, \\
 \bigl[\mathbf{C}^{(3)}_{12}\bigr]_{ij} & = &  \int d^{3}x \biggl\{ B^{(1)}_{i} \biggl[ \partial_{r} + \frac{\mu_{2}}{r} \biggr] B^{(2)}_{j} \biggr\}. \\
 \end{eqnarray}
 We also have
 \begin{eqnarray}
 \label{eq:explicit_nRR2}
 \bigl[\mathbf{S}^{(1)}_{11}\bigr]_{ij} & = &  \int d^{3}x \biggl\{ B^{(1)}_{i}B^{(1)}_{j} \biggr\}  = \bigl[\mathbf{S}^{(2)}_{11}\bigr]_{ij},\\
 \bigl[\mathbf{S}^{(1)}_{22}\bigr]_{ij} & = &  \int d^{3}x \biggl\{ B^{(2)}_{i}B^{(2)}_{j} \biggr\} = \bigl[\mathbf{S}^{(2)}_{22}\bigr]_{ij} .\\
 \end{eqnarray}
 The last expressions can then be expressed in prolate spheroidal coordinates with (\ref{eq:deri_r}),(\ref{eq:deri_z}) and (\ref{eq:meas_pro_coord}).

\section{Explicit expression for the Rayleigh-Ritz method with KBBF }
\label{app:explicit_RRKBBF}

The explicit expression of matrices $\mathbf{A}_{11},\mathbf{A}_{22}$ and $\mathbf{A}_{12}$ is again obtained by using the Dirac equation in cylindrical coordinates given in \cite{0305-4470-16-9-024}. Then, combining the ansatz in Eq. (\ref{eq:ansatz}) with the basis function expansion, we get
\begin{eqnarray}
\label{eq:explicit_nRR3}
\bigl[ \mathbf{C}^{(1)}_{11} \bigr]_{ij} & = &  \int d^{3}x \biggl\{ (V+mc^{2})B^{(1)}_{i}B^{(1)}_{j} \biggr\}, \\
\bigl[ \mathbf{C}^{(1)}_{22} \bigr]_{ij} & = &  \int d^{3}x \biggl\{ (V+mc^{2})B^{(2)}_{i}B^{(2)}_{j} \biggr\}, \\
\bigl[ \mathbf{C}^{(2)}_{11} \bigr]_{ij} & = &  \int d^{3}x \biggl\{ \biggl[  (\partial_{z}B^{(1)}_{i})(\partial_{z}B^{(1)}_{j}) 
+(\partial_{r}B^{(1)}_{i})(\partial_{r}B^{(1)}_{j}) \nonumber \\
&&\quad \quad \quad + \frac{\mu_{1}^{2}}{r^{2}} B^{(1)}_{i}B^{(1)}_{j}  - \frac{\mu_{1}}{r} B^{(1)}_{i}(\partial_{r}B^{(1)}_{j}) \nonumber \\
&&\quad \quad \quad - \frac{\mu_{1}}{r} (\partial_{r}B^{(1)}_{i})B^{(1)}_{j}  \biggr]  \frac{(V-mc^{2})}{4m^{2}c^{2}}\biggr\} ,\\
\bigl[ \mathbf{C}^{(2)}_{22} \bigr]_{ij} & = &   \int d^{3}x \biggl\{ \biggl[ (\partial_{z}B^{(2)}_{i})(\partial_{z}B^{(2)}_{j}) 
+(\partial_{r}B^{(2)}_{i})(\partial_{r}B^{(2)}_{j}) \nonumber \\
&&\quad \quad \quad + \frac{\mu_{2}^{2}}{r^{2}} B^{(2)}_{i}B^{(2)}_{j} + \frac{\mu_{2}}{r} B^{(2)}_{i}(\partial_{r}B^{(2)}_{j}) \nonumber \\
&&\quad \quad \quad + \frac{\mu_{2}}{r} (\partial_{r}B^{(2)}_{i})B^{(2)}_{j}  \biggr] \frac{(V-mc^{2})}{4m^{2}c^{2}} \biggr\},\\
\bigl[\mathbf{C}^{(2)}_{12}\bigr]_{ij} & = & \int d^{3}x \biggl\{ \biggl[ (\partial_{z}B^{(1)}_{i})(\partial_{r}B^{(2)}_{j}) 
+ \frac{\mu_{1}}{r} B^{(1)}_{i}(\partial_{z}B^{(2)}_{j})  \nonumber \\ 
&&\quad \quad \quad  -(\partial_{r}B^{(1)}_{i})(\partial_{z}B^{(2)}_{j})
+ \frac{\mu_{2}}{r} (\partial_{z}B^{(1)}_{i})B^{(2)}_{j} \biggr] \nonumber \\
&&\quad \quad \quad  \times \frac{(V-mc^{2})}{4m^{2}c^{2}} \biggr\},\\
\bigl[\mathbf{C}^{(3)}_{11}\bigr]_{ij} & = &  \int d^{3}x \biggl\{  (\partial_{z}B^{(1)}_{i})(\partial_{z}B^{(1)}_{j}) 
+(\partial_{r}B^{(1)}_{i})(\partial_{r}B^{(1)}_{j}) \nonumber \\
&&\quad \quad \quad + \frac{\mu_{1}^{2}}{r^{2}} B^{(1)}_{i}B^{(1)}_{j}  - \frac{\mu_{1}}{r} B^{(1)}_{i}(\partial_{r}B^{(1)}_{j}) \nonumber \\
&&\quad \quad \quad - \frac{\mu_{1}}{r} (\partial_{r}B^{(1)}_{i})B^{(1)}_{j}  \biggr\}\frac{1}{2m}  ,\\
\bigl[\mathbf{C}^{(3)}_{22}\bigr]_{ij} & = &   \int d^{3}x \biggl\{  (\partial_{z}B^{(2)}_{i})(\partial_{z}B^{(2)}_{j}) 
+(\partial_{r}B^{(2)}_{i})(\partial_{r}B^{(2)}_{j}) \nonumber \\
&&\quad \quad \quad + \frac{\mu_{2}^{2}}{r^{2}} B^{(2)}_{i}B^{(2)}_{j} + \frac{\mu_{2}}{r} B^{(2)}_{i}(\partial_{r}B^{(2)}_{j}) \nonumber \\
&&\quad \quad \quad + \frac{\mu_{2}}{r} (\partial_{r}B^{(2)}_{i})B^{(2)}_{j}  \biggr\} \frac{1}{2m}  ,\\
\bigl[\mathbf{C}^{(3)}_{12}\bigr]_{ij} & = & \int d^{3}x \biggl\{  (\partial_{z}B^{(1)}_{i})(\partial_{r}B^{(2)}_{j}) 
+ \frac{\mu_{1}}{r} B^{(1)}_{i}(\partial_{z}B^{(2)}_{j})  \nonumber \\ 
&&\quad \quad \quad  -(\partial_{r}B^{(1)}_{i})(\partial_{z}B^{(2)}_{j})
+ \frac{\mu_{2}}{r} (\partial_{z}B^{(1)}_{i})B^{(2)}_{j} \biggr\}  \frac{1}{2m}.
\end{eqnarray}
We also have $\bigl[\mathbf{C}^{(3)}_{21}\bigr]_{ij} =  \bigl[\mathbf{C}^{(3)}_{12}\bigr]_{ji}$, $\bigl[\mathbf{S}^{(3)}_{11}\bigr]_{ij} = 2m \bigl[\mathbf{C}^{(3)}_{11}\bigr]_{ij}$, $\bigl[\mathbf{S}^{(3)}_{22}\bigr]_{ij} = 2m \bigl[\mathbf{C}^{(3)}_{22}\bigr]_{ij}$ and $\bigl[\mathbf{S}^{(3)}_{12}\bigr]_{ij} = 2m \bigl[\mathbf{C}^{(3)}_{12}\bigr]_{ij}$. The last expressions can then be expressed in prolate spheroidal coordinates with (\ref{eq:deri_r}),(\ref{eq:deri_z}) and (\ref{eq:meas_pro_coord}).

\begin{acknowledgments}
Funding of this research has been provided by the Centre de Recherches Math\'ematiques (CRM) and Canada Research Chair (CRC) program. Supercomputer facilities were made available by RQCHP. One of the authors would also like to thank Huizhong Lu for his help with computer related issues.
\end{acknowledgments}

\bibliography{bibliography}

\end{document}